\documentclass{elsart}
\usepackage{epsfig}
\usepackage[dvips]{color}
\newcommand{\btbl}{\begin{tabular}}
\newcommand{\etbl}{\end{tabular}}
\newcommand{\beq}{\begin{displaymath}}
\newcommand{\eeq}{\end{displaymath}}
\newcommand{\ben}{\begin{equation}}
\newcommand{\een}{\end{equation}}
\newcommand{\bea}{\begin{eqnarray}}
\newcommand{\eea}{\end{eqnarray}}
\def\bar{\begin{array}}
\def\ear{\end{array}}

\def\bra{\langle }
\def\ket{\rangle}

\begin{document}
\begin{frontmatter}
\title{Role of the initial conditions on the enhancement
of the escape time in static and fluctuating potentials}
\author{A.Fiasconaro\corauthref{cor}},
\corauth[cor]{Corresponding author}
\ead{afiasconaro@gip.dft.unipa.it}
\author{D.Valenti, B.Spagnolo}
\address{INFM and Dipartimento di
Fisica e Tecnologie Relative, Viale delle Scienze - 90128 Palermo,
Italy}

\begin{abstract}
 We present a study of
the noise driven escape of an overdamped Brownian particle moving
in a cubic potential profile with a metastable state. We analyze
the role of the initial conditions of the particle on the
enhancement of the average escape time as a function of the noise
intensity for fixed and fluctuating potentials. We observe the
noise enhanced stability effect for all the initial unstable
states investigated. For a fixed potential we find a peculiar
initial condition $x_c$ which separates the set of the initial
unstable states in two regions: those which give rise to
divergences from those which show nonmonotonic behavior of the
average escape time. For fluctuating potential at this particular
initial condition and for low noise intensity we find large
fluctuations of the average escape time.
\end{abstract}
\end{frontmatter}

\section{Introduction}
In the two last decades a remarkable increase has been done in the
study of noise induced effects on nonlinear nonequilibrium
systems, with attention also to complex and biological systems
\cite{Fre,Par,Gam,Tal}. The introduction of noise and of a
deterministic driving force in such a systems gives rise to
counterintuitive effects which explain apparently anomalous
behaviors in experiments. Nonlinear relaxation decay of physical
systems from an initial unstable or metastable state involves
fundamental aspects of non-equilibrium statistical mechanics.
Examples of these resonance-like phenomena are stochastic
resonance (SR) \cite{Gam,Man1}, resonant activation (RA)
\cite{Doe}, noise enhanced stability (NES) \cite{Man2,Agu}, noise
induced phase transitions (see ref.\cite{Fre}), etc... A
nonmonotonic behavior of the average escape time as a function of
the noise intensity was revealed in a numerical study of a
Brownian particle moving in a periodic fluctuating cubic potential
\cite{Day}.  This is the NES phenomenon: the stability of an
otherwise unstable system can be enhanced by the presence of a
finite amount of noise. The nonmonotonic behavior that often
occurs in NES effect contradicts standard Kramers-like behavior,
i.e. an exponential or monotonic decrease of the mean escape time
with noise intensity \cite{Kra}. The NES phenomenon was
experimentally detected in the transient dynamics of an unstable
physical system \cite{Man2} and observed in different physical
systems \cite{Hir}. Recently some works have been done studying
piecewise linear  potentials with a metastable state, and exact
evaluation of the decay time was obtained \cite{Agu}. In this work
we present a study of the average decay time of an overdamped
Brownian particle subject to a cubic potential with a metastable
state at different unstable initial conditions. A relevant result
of this work is the existence, for fixed potential, of a peculiar
initial position $x_c$ of the Brownian particle, which corresponds
to the intersection point between the potential profile and the
$x$-axis (see Fig. 1). For all the initial unstable states between
the maximum of the potential and $x_c$, the average escape time
diverges, while for initial positions $x_o
> x_c$ we find a nonmonotonic behavior of the same quantity. Our
results are consistent with that obtained in the case of piecewise
linear potential profile \cite{Agu}. For a periodic fluctuating
potential we recover the NES phenomenon for different values of
the amplitude and of the frequency of the periodical driving force
and large fluctuations of the average escape time as a function of
the amplitude and the frequency at $x_o = x_c$.
\begin{figure}[htbp]
 \begin{center}
  \includegraphics[height=7cm]{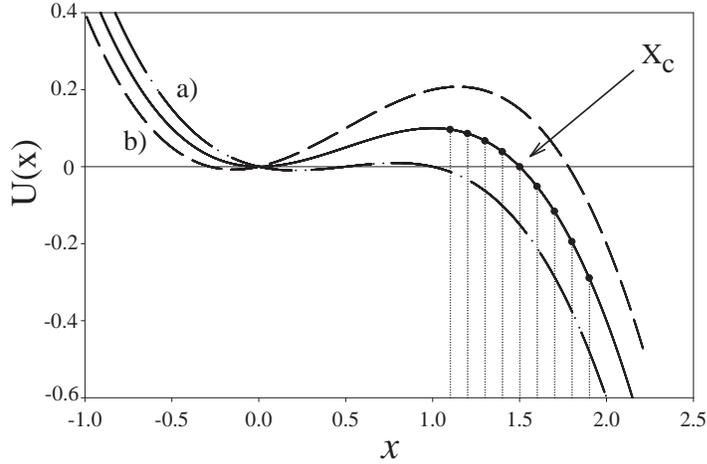}
  \linespread{1.2} 
  \caption{The cubic potential $U(x)$ with the various initial
  positions investigated (dots); $x_c$ is the critical initial position.
  (a) and (b) indicate the limit curves of
  the oscillating potential. The absorbing boundary is $x_F = 20$.}
  \label{fpot}
 \end{center}%
\end{figure}
\section{The Model}
 The starting point of our study is the following Langevin
 equation
 \ben
 \label{eq}
  \dot{x}=-\frac{\partial U(x,t)}{\partial x} + \sqrt{D} \xi(t)
 \een
 where $\xi(t)$ is the white Gaussian noise with the usual statistical properties:
 $\bra \xi(t) \ket = 0$ and $\bra \xi(t)\xi(t+\tau) \ket = \delta
 (\tau)$, and

 \ben
 \label{u1}
  U(x,t)=ax^2 - bx^3 - xAcos(\omega t),
 \een
 is the potential shown in Fig. 1, with  $A$ and $\omega$
 respectively the amplitude and the
frequency of the driving force, and $a=0.3$, $b=0.2$. The
potential profile has a local stable state
 at $x_0=0$  and an unstable state at $x_0=1$.
For fixed potential $A = 0$ the average escape time of a particle
starting from $x_o$ and reaching a final position $x_F$ is given
by \cite{Gar}

 \ben
 \label{tau}
  \tau(x_0,x_F)=\frac{2}{D} \int_{x_0}^{x_F} e^{2u(z)}
  \int_{-\infty}^{z} e^{-2u(y)} \;\; dy dz.
 \een
 where $u(y) = (0.3y^2 - 0.2y^3)/D$ is a dimensionless potential profile.
 We evaluate this double integral partly analytically and
partly numerically, splitting it as

 \ben
  \tau(x_0,x_F)=\frac{2}{D} \int_{x_0}^{x_F} e^{2u(z)}dz
  \left[\int_{-\infty}^{0} e^{-2u(y)} + \int_{0}^{z} e^{-2u(y)}\; dy \right],
  \label{tau1}
  \een
and evaluating analytically the first term inside the square
parenthesis:

\ben
 \int_{-\infty}^{0} e^{-2u(y)} =
 0.6046 \; e^{-z} [I_{-1/3}(z)+I_{1/3}(z)]
 -\frac{1}{2} \; {}_{2}F_{2}({\frac{1}{2},1};{\frac{2}{3},\frac{4}{3}};-2z),
   \label{tau2}
 \een
where $z = 1/(10D)$, $I_n(z)$ is the modified Bessel function of
the first kind and $_{p}F_{q}(a_1,a_2;b_1,b_2;z)$ is the
generalized hypergeometric function.
\begin{figure}[htbp]
 \centering
 \includegraphics[height=8cm]{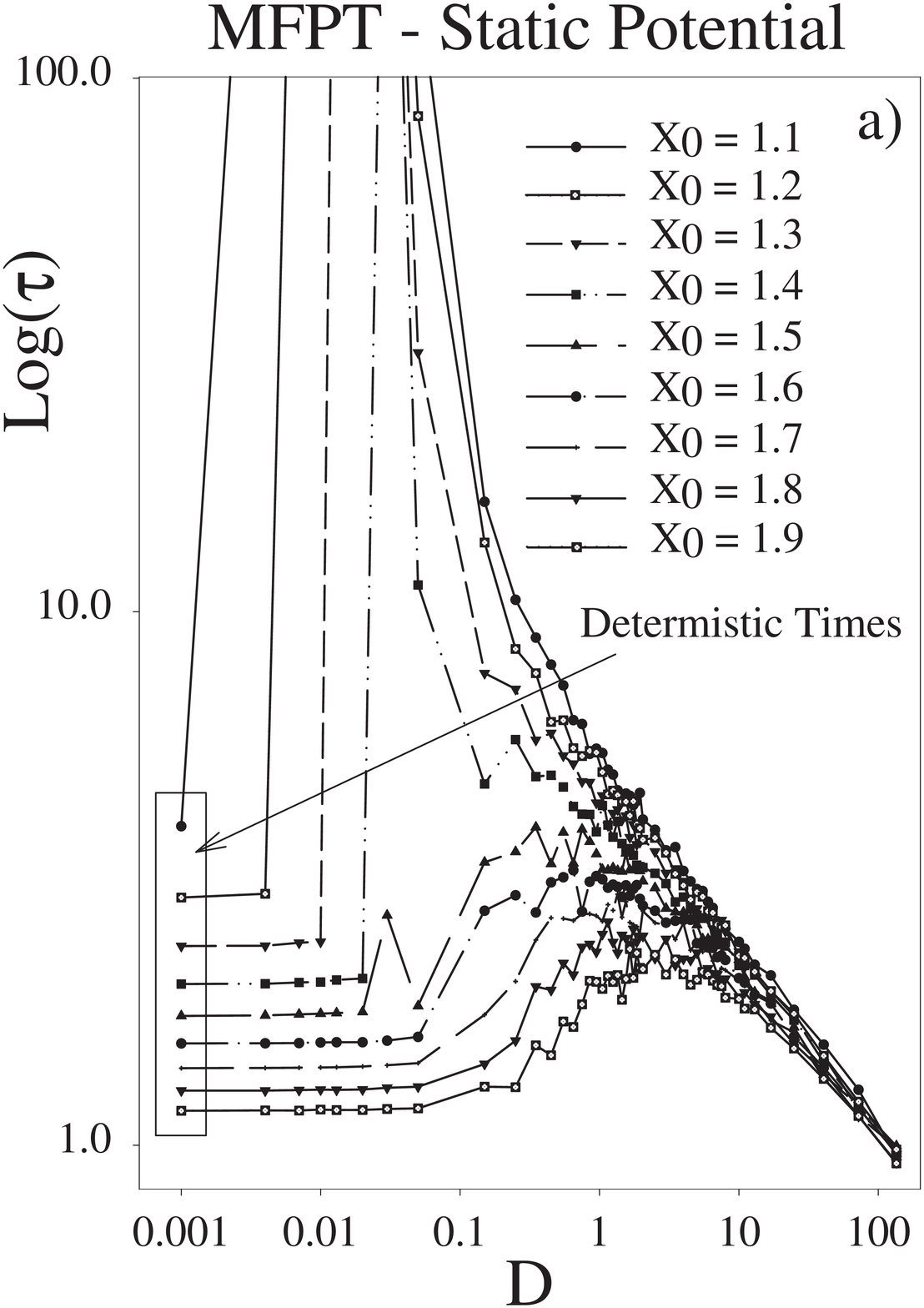}
 \includegraphics[height=8cm]{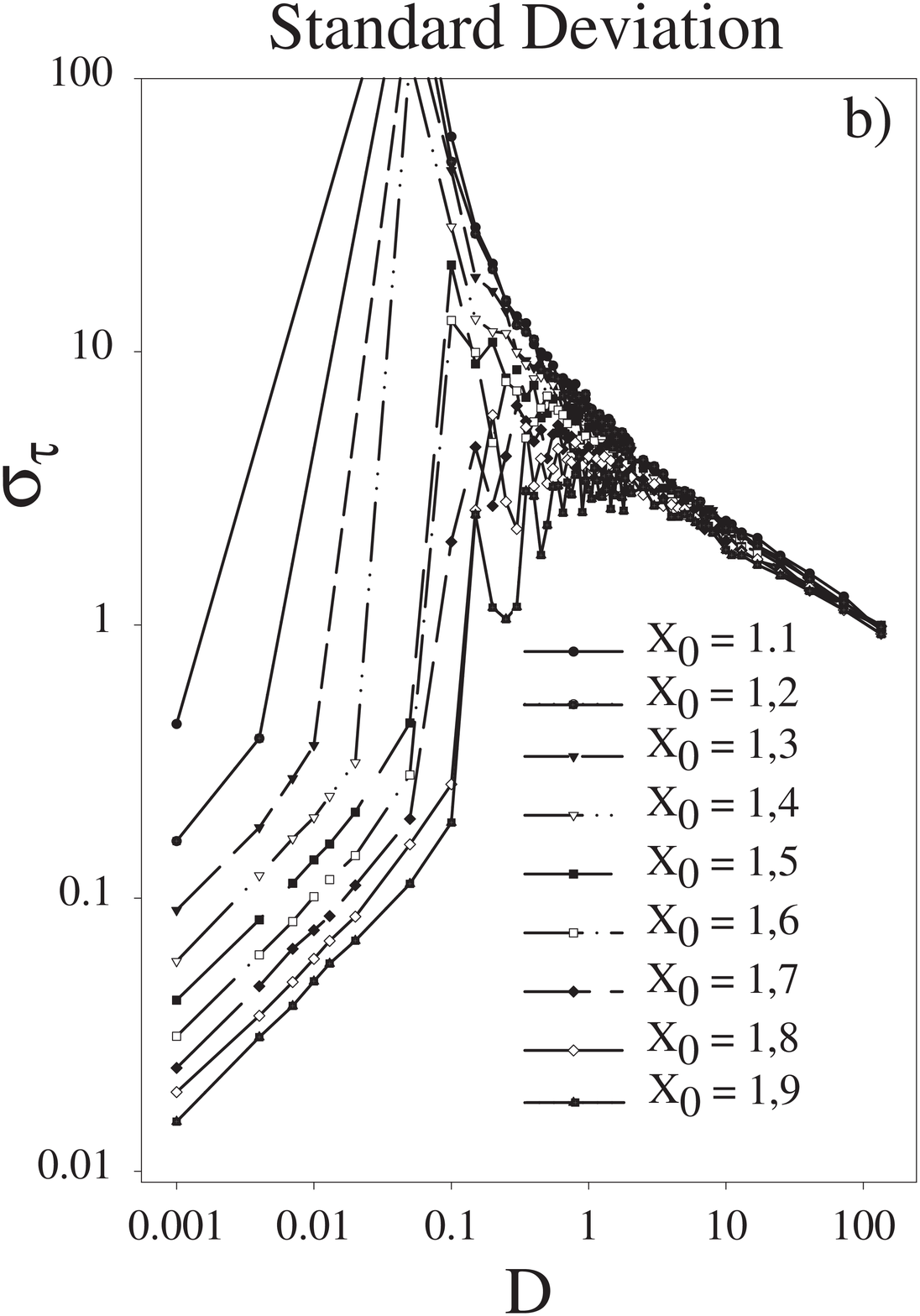}
 \linespread{1.2} 
 \vskip -0.5cm
 \caption{a) Mean First Passage Time as a function of noise
intensity for all the nine initial positions investigated (see
Fig. \ref{fpot}), namely: $x_o=1.1 \div 1.9$, with steps of $0.1$.
The number of realizations is 2000 and $x_F = 20$. b) The standard
deviation of the first passage time distribution, which
 shows a nonmonotonic behavior similar to that of MFPT.
 }
 \label{fnes}
\end{figure}

\section{Results and Comments}
\begin{figure}[htbp]
\begin{center}
 \includegraphics[height=7.cm]{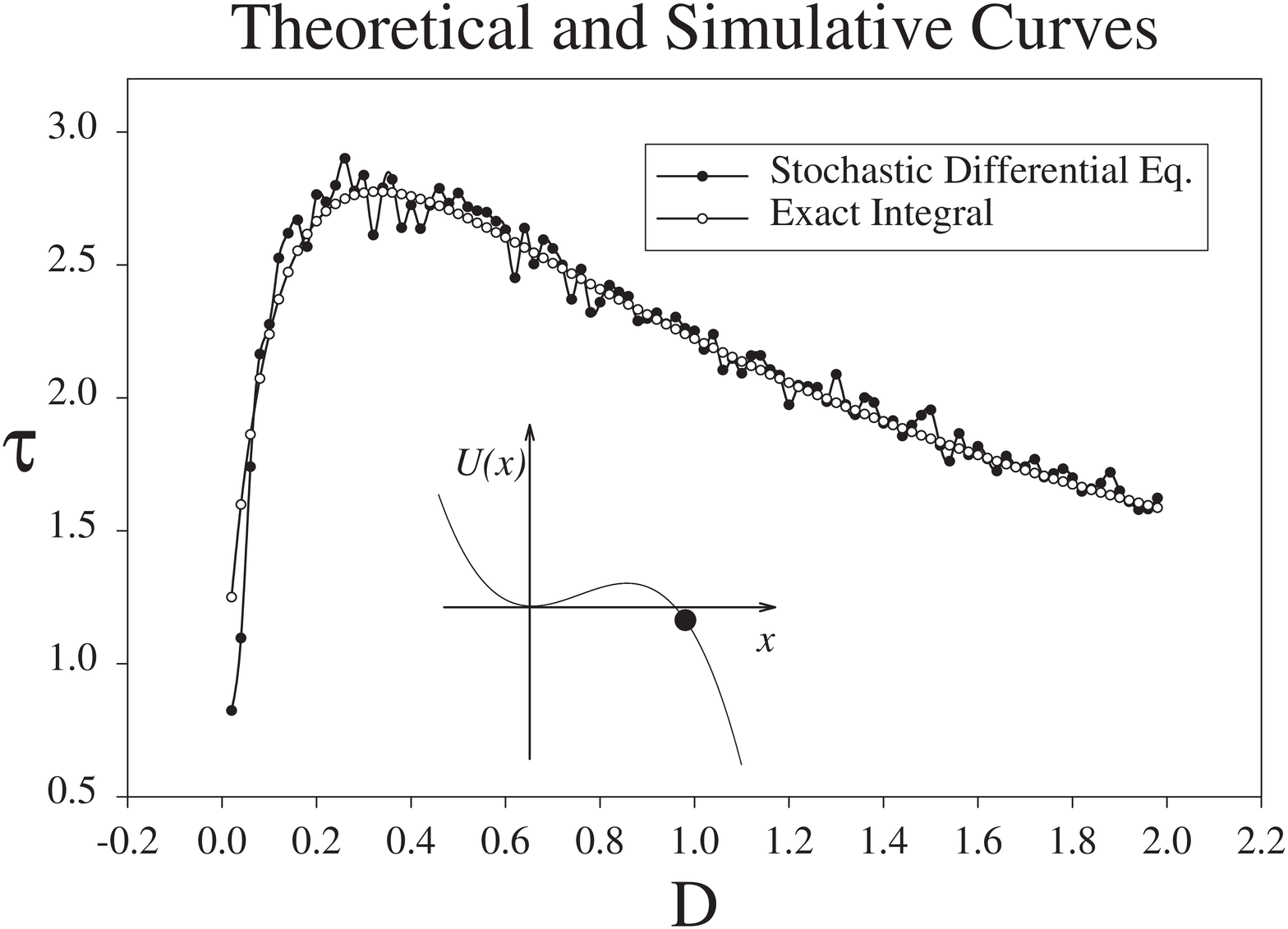}
 \linespread{1.2} 
 \vskip -0.5cm
  \caption{Average escape time as a function of the noise intensity $D$,
 for an initial position $x_o = 1.51 > x_c=1.5$ and $x_F = 2.2$.
 Comparison between theoretical MFPT (full circles)
 and the simulation  of the SDE (empty circles)(Eq. (\ref{eq})).}
 \label{ftheor}
\end{center}
\end{figure}

For all the initial unstable states beyond the potential barrier
(see Fig. \ref{fpot}) we find an enhancement of the average escape
time with respect to the deterministic time as a function of the
noise intensity (Fig. \ref{fnes}). Specifically, for high values
of the noise intensity with respect to the height of the barrier,
we recover the Kramers behavior. For intermediate values of noise
intensity, we observe an increase of the MFPT whose maximum goes
to infinity when the initial position is in the range $x_{max}<
x_o < x_c$. These results are quite consistent with those obtained
in the case of piecewise linear potential profiles \cite{Agu},
where a divergent behavior for $D \rightarrow 0$ is found. On the
other hand for very low values of noise intensity we obtain the
deterministic decay time. This effect, which appears in
contradiction with the divergence for $D \rightarrow 0$, can be
explained with the finite ensemble of particles in numerical
experiments. Indeed, for low noise intensity the probability that
some particle is trapped is very low and decreases exponentially
to zero, that is only very few particles will be trapped. Only
these particles, which represent very rare events for very small
noise values, contribute to the enhancement of the escape time and
give rise to the divergence. Therefore, because of this, we do not
observe such particles in simulation and the average escape time
becomes equal to the deterministic time. In Fig. \ref{fnes} we can
clearly see this effect. For initial positions approaching the
maximum of the potential this effect is less pronounced, because
the trapping probability of the particle is proportional to
$exp(-\Delta u(x_o)/D)$, where $\Delta u(x_o)$ is the potential
barrier ``seen" from the particle at $x(0)=x_o$. When the particle
is near the maximum a relatively low amount of noise can push the
particle back into the ``stable" state, where it remains trapped
for a long time, because of the low noise intensity.
 For higher values
of initial position the amount of noise to put the particle into
the potential well must be higher. This explains why the maximum
of the average decay time is shifted towards higher values of the
noise intensity. The observation time in digital simulation is
finite, therefore the simulated escape time takes the maximum at
this limiting point, and this is what we find (see Fig.
\ref{fnes}a). For all the initial conditions within the range:
$x_m < x_o < x_c$, we find that the average escape time can be
more and more increased by the noise, while for initial conditions
$x_o > x_c$ we obtain nonmonotonic behavior with a maximum and a
finite average escape time when the noise intensity goes to zero.
We calculate also the standard deviation ($\sigma_\tau$) of MFPT
as a function of the noise intensity for all the initial positions
investigated (see Fig.\ref{fnes}b). For $x_o > x_c$ we obtain a
nonmonotonic behavior like the MFPT behavior. For $x_o < x_c$ the
standard deviation grows towards divergent values for $D
\rightarrow \infty$, and this corresponds to a big tail in the
first passage time distribution. From the above analysis it
appears that a peculiar initial position for the potential profile
of Eq.(\ref{u1}) is $x_o=x_c=1.5$. We calculate then the
theoretical average escape time from Eqs.(\ref{tau1},\ref{tau2})
for an initial position just after the cross point $x_c$
 and we show the result in Fig. \ref{ftheor}. The agreement
 between the numerical simulation of the SDE of
the Eq. (\ref{eq}) and the theoretical evaluation is very good.

\begin{figure}[htbp]
 \begin{minipage}[t]{0.45\linewidth}
  \centering
  \includegraphics[height=10.cm]{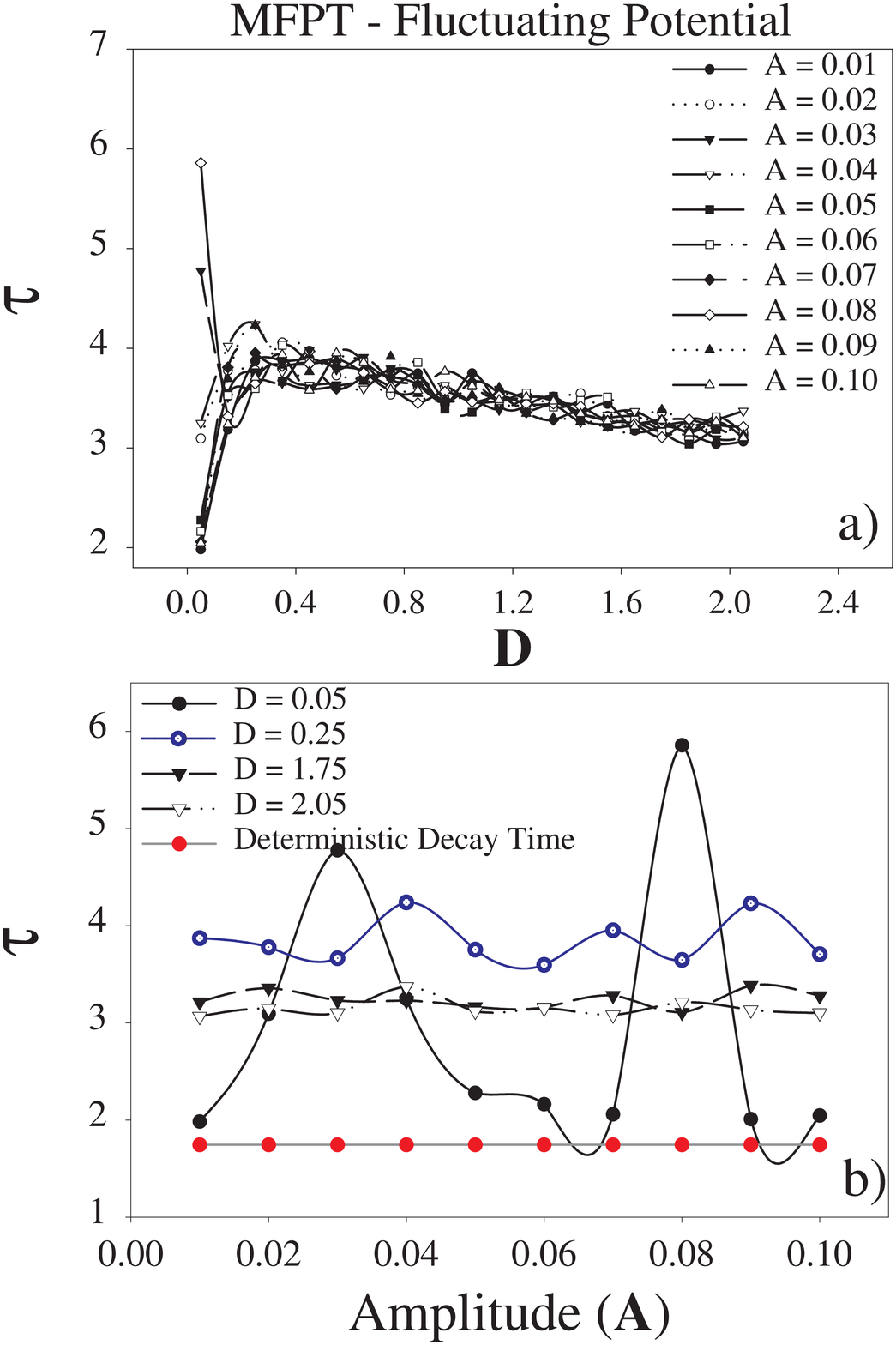}
  \linespread{1.2} 
  \caption{MFPT as a function of noise intensity (a) and of the amplitude
(b) of the driving force. The parameter values are: $x_o = 1.5,
\nu = 10 Hz$.}
  \label{foscam}
 \end{minipage}%
 \hfill\begin{minipage}[t]{0.45\linewidth}
  \centering
  \includegraphics[height=10.cm]{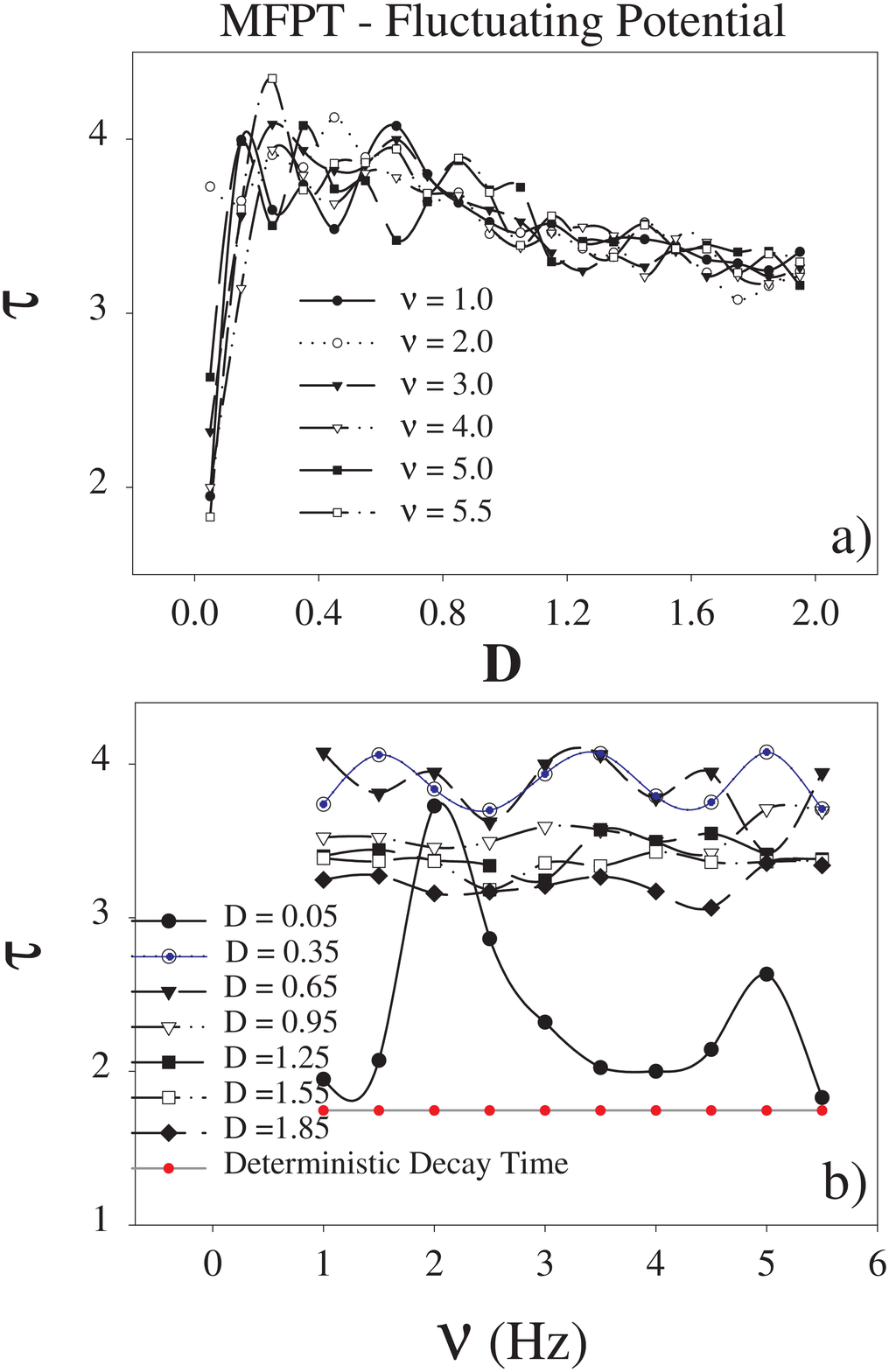}
  \linespread{1.2} 
  \caption{MFPT as a function of noise intensity (a) and of the frequency
(b) of the driving force. The parameter values are: $x_o = 1.5, A
= 0.02$.}
  \label{foscom}
 \end{minipage}
\end{figure}

We calculate the average escape time at the static cross position
$x_c=1.5$ for a periodical fluctuating cubic potential and the
results are shown in Fig.\ref{foscam} and Fig.\ref{foscom}.
 Here we recover
the enhancement of stability of the metastable state for all
values of the amplitude and of the frequency of the periodical
driving force investigated, with a nonmonotonic behavior of MFPT
(a-figures). We find also that (b-figures): (i) the MFPT values
are almost independent from the amplitudes and the frequencies of
the driving force, except for the case $D=0.05$; (ii) very large
fluctuations of MFPT as a function of both the amplitude and the
frequency of the periodic driving force at $D=0.05$ are present.
The origin of these large fluctuations can be ascribed to the
particular initial position represented by $x_c$, which is one of
the two boundaries delimiting the parameter region (A,$\nu$) where
NES effect can be observed \cite{Agu}. Because of the oscillation
of the potential between two limiting curves the intersection
point is a function of the time $x_c(t)$ and as a consequence the
Brownian particle experiences, during each period of the driving
force, the two different dynamical regimes of the case of the
static potential discussed above for $D \rightarrow 0$.
 Specifically the escape time
passes from a divergent regime ($x_c(t) < x_c$) to a convergent
one ($x_c(t) > x_c$), where $x_c$ is the intersection point for
the static potential. This effect becomes more important for low
noise intensity, i.e. for noise intensities less than the barrier
height, as we can see in Figs.\ref{foscam} and \ref{foscom} for
$D=0.05$. We obtain the same large fluctuations for other values
of $D$ less than the barrier height (0.1), not reported here for
sake of clearness of the figures. For values of amplitude and
frequencies of the driving force higher then those presented in
the figures, up to $\nu=20$ and up to $A=3$ respectively, the
calculations confirm these large fluctuations. We also
investigated the influence of different initial phases on the MFPT
behavior as a function of the frequency and we recover large
fluctuations again for $D=0.05$.
\section{Conclusions}
Nonmonotonic behavior of the mean escape time as a function of
noise intensity is a noise-induced effect for nonlinear
nonequilibrium systems with metastable states. In this work we
analyzed the role of the initial conditions on the enhancement of
the escape time from initial unstable states for a cubic
potential. We obtain NES effect for static and periodical
fluctuating potential and an enhancement of the NES effect for
initial positions between the maximum of the potential well and
the cross point $x_c$. We find also large fluctuations of the
average escape time as a function of both the amplitude and the
frequency of the sinusoidal force for low noise intensity, due to
the different dynamical regimes experienced by the Brownian
particle. Our results obtained for a particle moving in a cubic
potential are quite general, because we always obtain NES effect
when a particle is initially located just to the right of a local
potential maximum, with a local minimum in its left side and the
global escape region in its right side.

 \section{Acknowledgement}
We thank Dr. A. La Barbera for useful discussion and suggestions
concerning the numerical simulations. This work was supported by
INFM, MIUR and by \mbox{INTAS Grant 01-0450}.

  \linespread{1.1} 
\bibliographystyle{amsplain}
\bibliography{xbib}

\end{document}